\documentclass[11pt]{article}
\usepackage{graphicx}
\usepackage{dcolumn}
\usepackage{bm}
\usepackage{amssymb}
\usepackage{amsmath}
\usepackage{graphicx}

\usepackage[T1]{fontenc}
\usepackage[latin9]{inputenc}
\usepackage{geometry}
\usepackage{textcomp}
\usepackage{relsize}
\def\bew{\begin{widetext}}    
\def\eew{\end{widetext}}      

\def\beq{\begin{eqnarray}}    
\def\eeq{\end{eqnarray}}      

\newcommand{\ORo}{\Omega_{r}^0}
\newcommand{\OBo}{\Omega_{b}^0}
\newcommand{\OL}{\Omega_{\Lambda}}

\newcommand{\OLo}{\Omega_{\Lambda}^0}
\newcommand{\OX}{\Omega_{X}}

\newcommand{\OD}{\Omega_{D}}

\newcommand{\OR}{\Omega_r}

\newcommand{\rLeff}{\rho_{\CC,{\rm
eff}}}
\newcommand{\rco}{\rho_{c}^0}

\newcommand{\rM}{\rho_m}

\newcommand{\rR}{\rho_r}
\newcommand{\rB}{\rho_b}
\newcommand{\rD}{\rho_D}

\newcommand{\rX}{\rho_X}
\newcommand{\rinv}{\rho_{\rm inv}}

\newcommand{\rL}{\rho_{\CC}}
\newcommand{\rLi}{\rho_{\CC}^i}
\newcommand{\pL}{p_{\CC}}
\newcommand{\rLo}{\rho_{\CC}^0}
\newcommand{\pD}{p_D}

\newcommand{\wL}{\omega_{\CC}}
\newcommand{\CC}{\Lambda}

\begin{document}
\title{Relaxing a large cosmological constant}

\author{%
\textbf{Florian Bauer}$^{1}$\footnote{fbauer@ecm.ub.es}, %
\textbf{Joan Sol\`a}$^{1}$\footnote{sola@ecm.ub.es}, %
\textbf{Hrvoje \v{S}tefan\v{c}i\'{c}}$^{2}$\footnote{shrvoje@thphys.irb.hr}%
\vspace*{0.5cm}
\\
\small $^{1}$ High Energy Physics Group, Dept. ECM, and Institut de Ci{\`e}ncies del Cosmos\\
\small Univ. de Barcelona, Av. Diagonal 647, E-08028 Barcelona, Catalonia, Spain\\
\small $^{2}$ Theoretical Physics Division, Rudjer Bo\v{s}kovi\'{c}
Institute, PO Box 180, HR-10002 Zagreb, Croatia
}

\date{}

\maketitle

\begin{abstract}
\vspace{0.5cm} The cosmological constant (CC) problem is the
biggest enigma of theoretical physics ever. In recent times, it
has been rephrased as the dark energy (DE) problem in order to
encompass a wider spectrum of possibilities. It is, in any case,
a polyhedric puzzle with many faces, including the cosmic
coincidence problem, i.e.\ why the density of matter $\rM$ is
presently so close to the CC density $\rL$. However, the oldest,
toughest and most intriguing face of this polyhedron is the big
CC problem, namely why the measured value of $\rL$ at present is
so small as compared to any typical density scale existing in
high energy physics, especially taking into account the many
phase transitions that our Universe has undergone since the early
times, including inflation. In this Letter, we propose to extend
the field equations of General Relativity by including a class of
invariant terms that automatically relax the value of the CC
irrespective of the initial size of the vacuum energy in the
early epochs. We show that, at late times, the Universe enters an
eternal de Sitter stage mimicking a tiny positive cosmological
constant. Thus, these models could be able to solve the big CC
problem without fine-tuning and have also a bearing on the cosmic
coincidence problem. Remarkably, they mimic the $\CC$CDM model to
a large extent, but they still leave some characteristic imprints
that should be testable in the next generation of experiments.

\end{abstract}

\begin{center}
Keywords: Cosmological constant, vacuum energy, relaxation, modified gravity.

PACS: 98.80.Es, 95.36.+x, 04.50.Kd
\end{center}



\vskip 6mm


\newpage

\section{Introduction}\label{introduction}
High Energy Physics is described by quantum field theory (QFT) and
string theory. Unfortunately, these theoretical descriptions are
plagued by large hierarchies of energy scales associated to the
existence of many possible vacua. Such situation is at the root of
the old and difficult CC problem \,\cite{weinberg89}, i.e.{}, the
formidable task of trying to understand the enormous ratio between
the theoretical computation of the vacuum energy density and its
observed value, $\rLo\sim 10^{-47}\,\text{GeV}^4$, obtained from
modern cosmological data\,\cite{cosmdata}. The extremal possibility
occurs when the Planck mass~$M_{P}\sim 10^{19}\,\text{GeV}$ is used
as the fundamental scale; then the ratio $M_P^4/\rLo$ becomes $\sim
10^{123}$. One may think that physics at the Planck scale is not
well under control and that this enormous ratio might be fictitious.
However, consider the more modest scale $v=2\,M_W/g\simeq
250\,\text{GeV}$ of the electroweak Standard Model (SM) of Particle
Physics (the experimentally most successful QFT known to date),
where $M_W$ and $g$ are the $W^{\pm}$ boson mass and $SU(2)$ gauge
coupling, respectively. In this case, that ratio reads $|\langle
V\rangle|/\rLo\gtrsim 10^{55}$, where $\langle V\rangle=
-(1/8)M_H^2\,v^2<0$ is the vacuum energy (i.e.\ the expectation
value of the Higgs potential) and $M_H\gtrsim 114.4\,\text{GeV}$ is
the lower bound on the Higgs boson mass. Although one may envisage
the possibility that there is a cancelation between the various
theoretical contributions to the physical CC (including the bare
value), this has never been considered a realistic option owing to
the enormous fine-tuning that it entails (which, in addition, must
be corrected order by order in perturbation theory).

In this Letter, we discuss a dynamical mechanism that protects the
Universe from any initial CC of arbitrary magnitude $|\rLi|\gg\rLo$,
which could emerge, for instance, from quantum zero-point energy
(contributing roughly $\sim m^4$ for any mass $m$), phase
transitions ($\rLi=\langle V\rangle$) or even vacuum energy at the
end of inflation. We admit that $\rL=\rL(t)$ (with $\rL(t_i)=\rLi$,
$\rL(t_0)=\rLo$) can actually be an effective quantity evolving with
time.

Phenomenological models with variable $\rL$ have been considered in
many places in the literature and from different perspectives, see
e.g.\ \cite{Overduin}. At the same time, models with variable CC with
a closer relation to fundamental aspects of QFT have also been
proposed\,\cite{Reuter,ShapiroSola,Babic,FBPhD}. In all these cases,
the effective quantity $\rL=\rL(t)$  still has an equation of state
(EOS) $\pL=-\rL$ and, in this sense, it can be called a CC term.

The basic framework of our proposal is the generalized class of
$\CC$XCDM models introduced in \cite{LXCDM}, in which there is a
fixed or variable $\rL$ term together with an additional
``effective'' component $X$ (in general \textit{not} related to a
fundamental, e.g.\ scalar, field). This particular class of variable
CC models is especially significant in that they could cure the
cosmic coincidence problem\,\cite{LXCDM} in full consistency with
cosmological perturbations\,\cite{GPS}. Here we present a
generalization of these models that might even cure the old
(``big'') CC problem\,\cite{weinberg89}. Recently,
in\,\cite{Stefancic08} a model along these lines was introduced with
a DE density~$\rD$ and an inhomogeneous EOS $\pD=\omega\,\rD-\beta
H^{-\alpha}$ which includes a term proportional to the negative
power of the Hubble rate~$H$. This additional term becomes
sufficiently large to compensate an initial $\rLi$ when this is
about to dominate the universe and forces it eventually into a final
de~Sitter era with a small CC. For recent related work on relaxation
mechanisms, see e.g.\ \cite{Barr:2006mp,Batra:2008cc,Diakonos:2007au}.
In a different vein, the CC problem can also be addressed in
quantum cosmology models of inflation, through the idea of
multiuniverses\,\cite{Linde87} and the application of anthropic
considerations\,\cite{weinberg89}.

Let us recall that, historically, most of the models addressing the
relaxation of the CC were based on dynamical adjustment mechanisms
involving scalar field potentials\,\cite{oldrelax}. In the present
work, the relaxation mechanism that we propose is also dynamical, it
does not require any fine-tuning and, as noted, it does not depend
in general on scalar fields. To be more precise, the model we
present here is a $\CC$XCDM relaxation model of the CC, which
includes also matter and radiation eras. We study the two
possibilities $\rLi<0$ and $\rLi>0$, with arbitrary value. For
$\rLi<0$, our scenario avoids the big crunch at early times and
allows the cosmos to evolve starting from a radiation regime with
subsequent matter and de Sitter eras like the standard $\CC$CDM
model.  Finally, let us emphasize that our method to tackle the CC
problem is formulated directly at the level of the (generalized)
field equations, rather than from an effective action functional. In
this sense, we follow the historical path of Einstein's derivation
of the original field equations. At the moment, a version
of our model with an action functional is not available, but its efficiency at the level of
the field equations is truly remarkable, as we will show. In this
sense, its phenomenological success may constitute a first
significant step in the way of finding a solution of the difficult
CC problem.

The present Letter is organized as follows. In section \ref{setup}
we present the basic setup of our model. In section \ref{toymodel}
we present a toy model of the CC relaxation mechanism which helps to
understand the basic idea behind our proposal, although it is still
too simple to describe our Universe. Only in section \ref{fullmodel}
we present a first realistic version of the full relaxation
mechanism and we perform a numerical analysis of it. In section
\ref{discussion}, we discuss in more detail some aspects and
implications of our model. Finally, in the last section we draw our
conclusions.

\section{The setup}\label{setup}

We start from the generalized Einstein field equations
\begin{equation}
R_{\mu \nu }-\frac{1}{2}g_{\mu \nu }R =
-\frac{8\pi}{M_P^2}\,(T_{\mu\nu}^m +
T^{X}_{\mu\nu}+g_{\mu\nu}\,\rLeff)\,,
\label{GEE}
\end{equation}
where $T_{\mu\nu}^m$ is the energy-momentum tensor of ordinary
matter -- including the energy densities of radiation ($\rR$) and
baryons ($\rB$). Furthermore, $T^{X}_{\mu\nu}$ describes the X
component ($\rX$), interacting with the effective CC term
$g_{\mu\nu}\,\rLeff$ in such a way that the total density of the
dark sector, $\rD=\rLeff+\rX$, is covariantly conserved (in
accordance with the Bianchi identity). The effective CC density
$\rLeff$ is given by
\begin{equation}
\rLeff=\rLi+\rinv\,.\label{eq:rhoL-Intro}
\end{equation}
Here, $\rLi$ is an arbitrarily large initial (and constant)
cosmological term, and $\rinv=\rinv(R,S,T)$ is some function of
the general coordinate invariant terms
\begin{eqnarray}
R & \equiv & R_{\mu\nu}g^{\mu\nu}=6H^{2}(1-q),\nonumber \\
S & \equiv & R_{\mu\nu}R^{\mu\nu}=12H^{4}\left[\left(\frac{1}{2}-q\right)^{2}+\frac{3}{4}\right],\label{eq:Invariants-RST}\\
T & \equiv & R_{\mu\nu\rho\sigma}R^{\mu\nu\rho\sigma}=12H^{4}(1+q^{2})\,,\nonumber
\end{eqnarray}
which we have evaluated in the flat Friedmann-Robertson-Walker (FRW)
metric in terms of the expansion rate~$H=\dot{a}/a$, and the
deceleration parameter
$q=-{\ddot{a}\,a}/{\dot{a}^2}=-{\dot{H}}/{H^{2}}-1$. We find it
useful to write the structure of $\rinv$ in the form
\begin{equation}
\rinv=\frac{\beta}{f}\,,\label{eq:rinv}
\end{equation}
where $\beta$ is a dimension $6$ parameter and $f=f(R,S,T)$ is a
dimension $2$ function of the aforementioned invariants. %
This form
is particularly convenient since the function $f$ must grow at high
energies and hence the vacuum energy is ultraviolet safe, i.e.\ in
the early Universe $\rinv\to 0$ and $\rLeff\to\rLi$, where $\rLi$ is
arbitrarily large but finite.

The generalized field equations (\ref{GEE}) fall into the
metric-based category of extensions of General Relativity.
However, at this point the following observation is in order. In
the literature, the extensions of Einstein's field equations are
usually of a restricted class, namely those that can be derived
from effective gravitational actions of the form
\begin{equation}\label{FRST}
\Gamma=\,\int d^4x\,\sqrt{-g}\,\left[\frac{M_P^2}{16\pi}\,R+
F(R,S,T)\right]\,.
\end{equation}
This class of models may be called the ``F(R,S,T)-theories'' as they
are characterized by an arbitrary (albeit sufficiently
differentiable) local function $F$ of the invariants defined in
Eqs.~(\ref{eq:Invariants-RST}), usually some polynomial of these
invariants. Work along these lines has been put forward e.g.\
in\,\cite{Carroll04}. The particular subclass of models in which the
function $F$ depends only on $R$, or ``F(R)-theories'', is
well known and has been subject of major
interest\,\cite{SotiriouFaraoni08,SaezElizalde09}.

However, as advertised in the introduction, in this work we
formulate the relaxation mechanism directly in terms of the
generalized field equations (\ref{GEE}), without investigating at
the moment the eventual connection with an appropriate effective
action. The reason is, basically, because we aim at maximal
simplicity at the moment. To be sure, after many years of
unsuccessful attempts, the CC problem has revealed itself as one
of the most difficult problems (if not the most difficult one) of
all theoretical physics; and we should not naively expect to shoot
squarely at it and hope to hit the jackpot at the first trial, so
to speak. In this sense, if we can find a way to solve, or at
least to significantly improve, the problem directly at the level
of the field equations, we might then find ourselves in a truly
vantage point to subsequently attempt solving the CC problem at
the level of some generalized form of the effective action of
gravity.

All in all, let us warn the reader that the connection between the
two approaches (viz.\ the one based on the field equations and the
functional one) is, if existent, non-trivial. In fact, we note
that the presumed action behind the field equations (\ref{GEE})
need not be of the local form (\ref{FRST}), and in general we
cannot exclude that it may involve some complicated contribution
from non-local terms. These terms, however involved they might
be, are nevertheless welcome and have been advocated in the
recent literature as a possible solution to the dark energy
problem from different perspectives \,\cite{DeserWoodard}. In the
present work, we wish to put aside the discussion of these terms,
and, for that matter, all issues related to the hypothetical
action functional behind our field equations. Instead, we want to
exclusively concentrate on the phenomenological possibilities
that our framework can provide on relaxing the effective CC term
(\ref{eq:rhoL-Intro}) depending on the choice of the function
$f$.  In particular, if the initial $\rLi$ is a very large
cosmological constant associated to a strong phase transition
(e.g.\ some GUT phase transition triggering the process of
inflation, the zero point energy of some field, or just the
electroweak vacuum energy of the SM), the late time behavior of
$\rLeff$ can be sufficiently tamed (without fine-tuning) so as to
be perfectly acceptable by the known cosmological data.

The main aim of the present approach is thus of practical nature; if
the CC problem can be efficiently tackled at the level of the field
equations to start with (something that, to the best of our
knowledge, has never been accomplished before), it should fit the
bill as it can be already a crucial first step in the path to solve
the CC problem -- namely, before unleashing a more formal (and,
predictably, even more difficult) theoretical assault to it at the
effective action level. We leave this part of the investigation for
future work, and we concentrate here on the potential
phenomenological benefits of assuming a set of generalized field
equations of the form (\ref{GEE}).

\section{Toy model}\label{toymodel}

Let us first illustrate the mechanism for a universe with only
radiation and with $f$ equal to just the Ricci scalar~$R$, so that
Eq.\,(\ref{eq:rhoL-Intro}) becomes
\begin{equation}
 \rLeff=\rLi+\frac{\beta}{6H^{2}(1-q)}.
\end{equation}
To start with, we consider the case $\rLi<0$ (as in the SM case) and
take $\beta>0$. Let us also assume a spatially flat Universe. At
late times, the de~Sitter regime is realized and the deceleration
approaches the value~$q\rightarrow-1$ while the Hubble rate becomes
constant and very small,~$H\rightarrow H_{*}$ (of order of the
current rate $H_0$). The vacuum energy density takes on the tiny
observed value $\rLeff\rightarrow\rho_{\CC}^{*}\simeq\rLo$, which is
related to~$H_{*}$ by the Friedmann equation:
$3M_{P}^{2}H_{*}^{2}=8\pi\rho_{\CC}^{*}$. Since the initial
value~$|\rLi |$ is assumed to be much larger than~$\rLo$, the final
Hubble rate is approximately given by
\begin{equation}\label{Hstar}
H_{*}^{2}\approx-\frac{\beta}{{12\rLi}}\,.
\end{equation}
We observe that the large $|\rLi|$ is responsible for the small
final value of~$H_{*}$, provided the parameter~$\beta$ has a
suitable order of magnitude and without any need of fine-tuning
because this relation does not include differences between large
numbers (cf.\ \cite{Quartin:2008px} for a  discussion on fine-tuning
issues). Moreover, this solution is stable. Indeed, the driving of
$H^2$ to small values by the large and negative $\rLi<0$ becomes
compensated by the positive second term in~$\rLeff$, which grows
as~$H$ decreases. On the other hand, a potential instability caused
by a growing $H$ would also be unharmful because it would make the
second term decrease, so that~$\rLi<0$ would stabilize $H$ again.

At earlier times, $H\gg H_{*}$ and the relaxation of
$\rho_{\Lambda}$ originates from the~$(1-q)$ factor in the
function~$f$, namely the negative~$\rLi$ drives dynamically the
deceleration parameter~$q$ to larger values until $q$ becomes very
close to~$1$, which corresponds to radiation-like expansion.
However, $q$ cannot cross $q=1$ from below since the (positive)
second term in~$\rLeff$ would dominate over~$\rLi$ and stop the
cosmic deceleration before $q$ reaches $1$. Summarizing, this
simplified model keeps the enormous vacuum energy~$\rLi$ under
control at any time thanks to the relaxation mechanism implemented
in the function~$f$ in Eq.~(\ref{eq:rinv}). Furthermore, it provides
a reasonable expansion history with radiation-like expansion
($q\simeq 1$, $H$ large) in the past and a stable de~Sitter solution
($q=-1$, $H=H_{*}\lesssim H_0$ tiny) in the future. The transition
is smooth and happens when the Hubble rate is sufficiently small to
ensure the CC relaxation as explained above.

\section{Full relaxation model}\label{fullmodel}

The simple model discussed above is able to handle the large
negative term~$\rLi$ without abrupt changes in the expansion
history. However, the model is unrealistic in that there is no
matter era yet, because $q$ will stay around the radiation
domination value~$q\simeq 1$ until the de Sitter phase starts.
Therefore, we have to make sure that the universe goes also through
the matter epoch by completing the structure of $\rLeff$ with a
term proportional to~$({1}/{2}-q)^{-1}$, which would work
like~$R^{-1}$ but with~$q={1}/{2}$ as the stabilizing point for the
next high $H$ interval. For this purpose, we use the scalar
invariants from Eq.~(\ref{eq:Invariants-RST}). A useful expression
is to involve not only $R$ but also $S$, as follows:
\begin{equation}
 R^{2}-S=24H^{4}(2-q)\,(1/2-q)\,.
\end{equation}
Notice that this combination is proportional to~$(1/2-q)$ and hence
allows the relaxation of the vacuum energy in the matter era. Again,
this expression alone would be unrealistic because it would enforce
the Universe to linger in the matter era and would prevent the
existence of a preceding radiation era. However, we can combine the
three invariants $R,S,T$ to finally form a much more realistic
ansatz for $f$ in Eq.\,(\ref{eq:rinv}), i.e.\ in such a way that the
radiation and matter epochs occur sequentially before the de~Sitter
universe is eventually reached in the infrared. Indeed, consider the
expression
\begin{eqnarray}\label{eq:rho_L}
f & = & \frac{R^{2}-S}{R}+y\cdot RT\\
 & = & 4H^{2}\frac{(\frac{1}{2}-q)(2-q)}{(1-q)}+y\cdot72H^{6}(1-q)(1+q^{2}).\nonumber \end{eqnarray}
We see that $f$ is constructed such that the first term contains
only two powers of~$H$, which ensures that the cosmic evolution is
reasonably close to that of the $\Lambda$CDM model during the matter
and subsequent de~Sitter stages.
\begin{figure}
\noindent \begin{centering}
\includegraphics[clip,width=1\columnwidth]{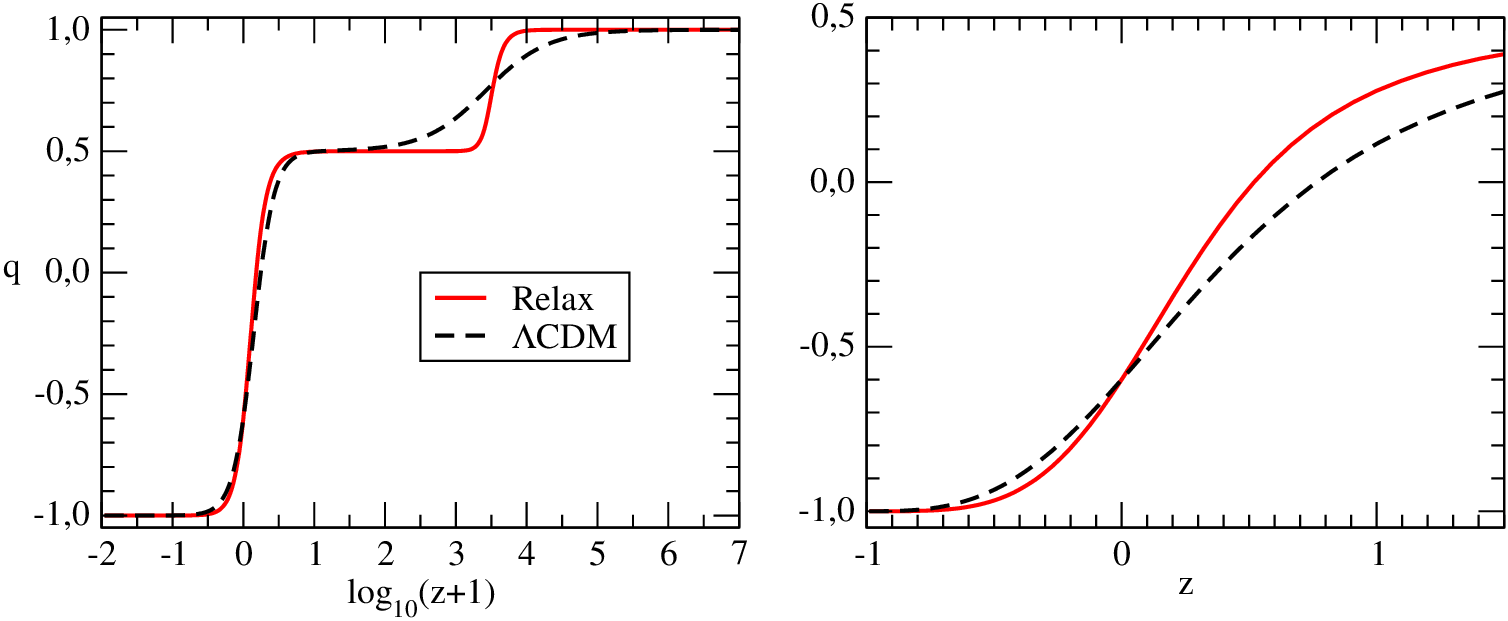}
\par\end{centering}
\caption{\label{fig:Decel-q} General behavior of the deceleration
parameter~$q$ in the relaxation CC model (\ref{eq:rhoL-Intro}) (solid) and in the $\Lambda$CDM model (dashed) as a
function of the cosmological redshift $z$. For any initial $\rLi$,
the universe goes through a radiation-dominated epoch ($q=1$), a
matter-dominated epoch ($q={1}/{2}$) and, eventually, into a final
de~Sitter phase ($q=-1$) with $\rLeff\simeq\rLo\ll|\rLi|$. Concrete
parameters in this plot are as in Fig.~\ref{fig:EnergyDensities}.}
\end{figure}
The second term contains~$T$ to provide a different scaling ($\sim
H^6$) in terms of the expansion rate. As a result, the
factor~$(1-q)$ will dominate over~$({1}/{2}-q)$ for large values of
$H$, i.e.{}, during the radiation regime.

The matter--radiation transition (``equality'') happens when both
terms in~$f$ are of the same magnitude and the corresponding time is
fixed by the parameter~$y\sim H_{eq}^{-4}$,
where~$H_{eq}\sim10^{5}H_{0}$ is the Hubble rate at equality. The
generic behavior of $q$ for the relaxation model under consideration
can be seen in Fig.~\ref{fig:Decel-q}. The transitions between
different epochs are not as smooth as in the $\Lambda$CDM cosmos,
although this feature depends on the detailed form of the
function~$f$. For our illustrative purposes, the qualitative results
given here should be sufficient to appreciate the virtues of this
relaxation mechanism. Its main benefit is the complete insensitivity
of the universe with respect to an initial cosmological constant
$\rLi$ of arbitrary magnitude (which in the present example we have
chosen to be negative).

Remarkably enough, our construction does not require to fine-tune
any of the parameters of the model. Thus, the insurmountable
problems associated to the traditional ``cancelation'' procedure can
be solved automatically through this dynamical relaxation mechanism,
which is triggered by pure gravitational physics (no scalar fields
at all). The current value~$\rLo$ of the effective vacuum energy and
the corresponding low Hubble rate~$H_{*}$ are fixed only by the
magnitude of the parameter~$\beta$, which is the $6^{\text{th}}$
power of a mass scale~$M$. Since $|\rLi|\gg\rLo$,
Eqs.~(\ref{eq:rinv}) and (\ref{eq:rho_L}) indicate that
\begin{equation} |\beta|\equiv M^{6}=|\rLi|\cdot f\sim |\rLi|
H_{*}^{2},\label{eq:beta}\end{equation}
where $H_{*}\simeq H_0\sim 10^{-42}\,\text{GeV}$. Remarkably, $M$
can be of the order of a typical Particle Physics scale; if e.g.\
the initial vacuum energy is $\sim M_P^4$, then $M\lesssim
100\,\text{ MeV}$ (i.e.{}, of the order of the characteristic QCD
scale $\Lambda_{QCD}$ where the lowest phase transition occurs in
the SM). The above relation (\ref{eq:beta}) can be rephrased in
another suggestive way. Since $q\simeq -1$ in the eventual de Sitter
regime -- which starts approximately near our time --, we find that
the current value of the CC (which is of the order of the asymptotic
value $\rLeff^{*}$) is roughly given by the appropriate ratio of the
two order parameters characterizing the most extreme phase
transitions ever occurred in our Universe:
\begin{equation}
\rLo\simeq \rLeff^{*}\simeq\frac{\Lambda_{QCD}^6}{100\,M_P^2}\,.
 \label{eq:beta2}\end{equation}
%


So far, the discussion of the CC relaxation was based only on the
form of $\rLeff$ in Eq.~(\ref{eq:rhoL-Intro}), and probably it can
be implemented in various ways without losing its benefits (e.g.\
models with inhomogeneous EOS, modified gravity Lagrangian). In the
following, we will discuss the concrete dynamics in a $\CC$XCDM-like
setup\,\cite{LXCDM}, where the total energy density $\rho_{\rm
tot}=\rR+\rB+\rD$ includes the usual components of the known
universe, $\rR$ and $\rB$ (i.e.\ radiation and baryons) as well as
the extra contributions from the dark sector: $\rD=\rX+\rLeff$. 
The conventional components are covariantly conserved leading to the usual scaling laws $\rho_{r}\propto a^{-4}$ and $\rho_{b}\propto
a^{-3}$. Since the dark sector does not interact with the conventional components, $\rD$ is conserved, too. %
From the Bianchi
identity satisfied by the terms on the \textit{l.h.s.}\ of
Eq.\,(\ref{GEE}) and the covariant conservation law of ordinary
matter ($\nabla^{\mu} T_{\mu\nu}^m=0$), the corresponding covariant
conservation in the dark sector reads
\begin{equation}\label{covariantlaws}
\nabla^{\mu}\left[T^{X}_{\mu\nu}+g_{\mu\nu}\,\rLeff\right]=0\,.
\end{equation}
Let us assume that $X$ is a pressureless component. Computing the
previous expression in the FLRW metric, it boils down to
\begin{equation}
\dot{\rho}_{\Lambda, {\rm
eff}}+\dot{\rho}_{X}+3H\rho_{X}=0\,.\label{eq:Bianchi}\end{equation}
This equation shows that the two components of $\rD$ are actually
interacting. The fact that the EOS parameter of $X$ is taken to be
$\omega_X=0$ (i.e.\ pressureless) is because $X$ can then mimic (and
can be referred to as) dark matter (DM). In this sense, the energy
density of the dark sector, $\rD$, can be thought of as being the
sum of the DM and DE (interacting) energy densities, where the DE
one is, in turn, the sum of the true cosmological constant $\rLi$
and the effective gravitational component $\rinv$, i.e.\
Eq.\,(\ref{eq:rhoL-Intro}). Thus, we have all the necessary
ingredients to implement realistically our Universe within the
relaxation model.

The basic dynamical equations read:
\begin{eqnarray}
H^{2}&=&\left(\frac{\dot{a}}{a}\right)^{2}={H_0^2}\,\frac{\rho_{\rm tot}}{\rco}=\frac{H_0^2}{\rco}(\rR+\rB+\rX+\rLeff)\,,\label{eq:Fried1}\\
qH^{2}&=&-\frac{\ddot{a}}{a}=\frac12\,\frac{H_0^2}{\rco}\sum_{n}\rho_{n}(1+3\omega_{n})=\frac{H_0^2}{\rco}\left(\rR+\frac{1}{2}\rB+\frac{1}{2}\rX-\rLeff\right)\,,\label{eq:Fried2}\end{eqnarray}
where $\rco\equiv \rho_{\rm tot}(t_0)=3H_0^{2}M_P^{2}/(8\pi)$ is the
current critical energy density.
\begin{figure}
\noindent \begin{centering}
\includegraphics[clip,width=1\columnwidth]{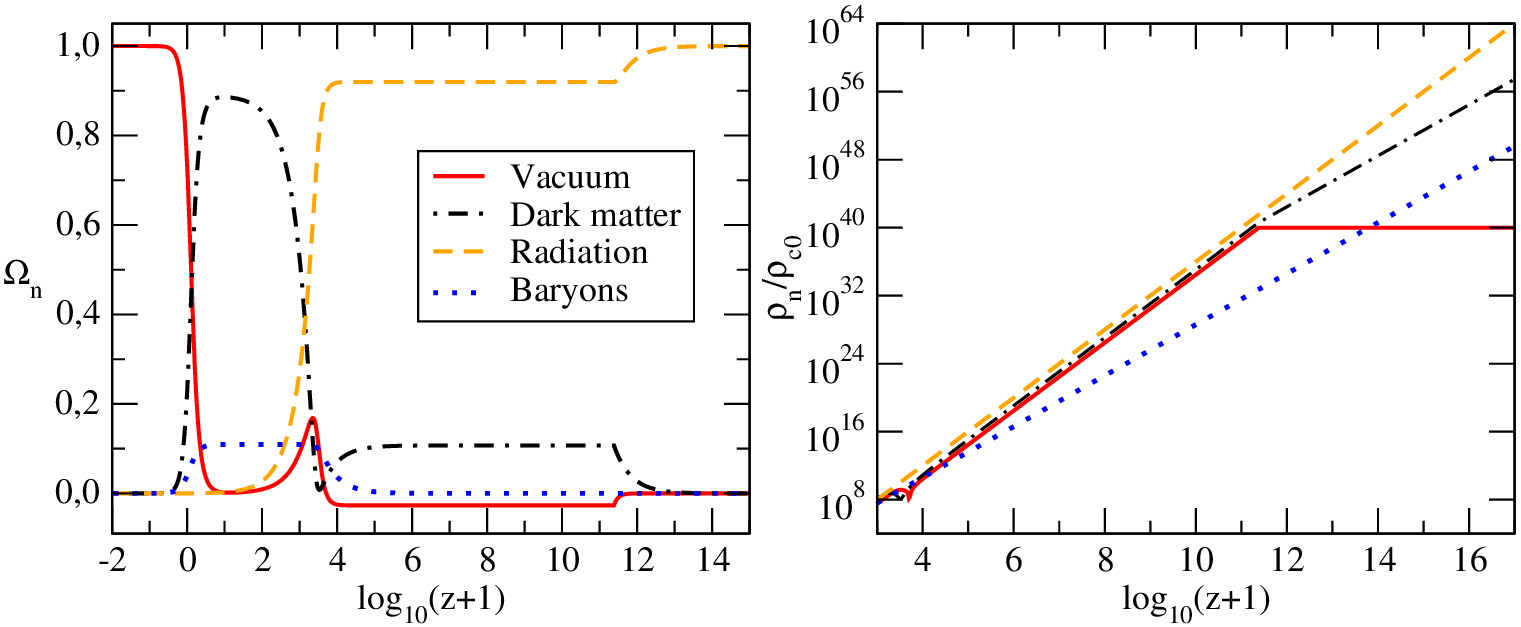}
\par\end{centering}
\caption{\label{fig:EnergyDensities}\label{fig:EnergyDensities-highz}\textit{Left}: Normalized energy
densities~$\Omega_{n}(z)=\rho_{n}(z)/\rho_{\rm tot}(z)$ for the CC,
dark matter, radiation and baryons. The initial CC is
$\rLi=-10^{40}\rco$. Parameters at $z=0$ read: $\OLo=0.73$,
$\ORo=10^{-4}$, $\OBo=0.04$, $q_0=-0.6$. The eventual de Sitter
regime is $\rho_{\rm tot}\to\rho_{\CC}^{*}\simeq\rco$. \textit{Right}: Absolute energy
densities~$\rho_{n}/\rco$ of the CC, dark matter,
radiation and baryons. Note that~$\rLeff<0$
and the plot shows~$|\rLeff|$.}
\end{figure}

The various EOS parameters for $n=r,b,X,\CC$ are
$\omega_{r}={1}/{3}$ (accounting for photons and light neutrinos)
and  $\omega_{b}=\omega_{X}=0$, $\wL=-1$. Using these equations, in
Fig.\,\ref{fig:Decel-q} we plot the numerical solution for $q(z)$
and in Fig.\,\ref{fig:EnergyDensities} we solve for the normalized
densities $\Omega_{n}(z)\equiv\rho_{n}(z)/\rho_{\text{tot}}(z)$,
where we have assumed an initial cosmological constant
$\rLi=-10^{40}\rco$. This should suffice to illustrate the great
efficiency of this relaxation mechanism. Since $y\simeq 10^{21}\,
H_{0}^{-4}$, Eq.~(\ref{eq:beta}) yields
$\beta\sim(10^{-3}\,\text{eV})^{6}$. Notice that in the presence of
several phase transitions, the value of $\beta\equiv M^6$ is fixed
by the strongest one. We have seen above that $M\lesssim
\Lambda_{QCD}\simeq 100\,\text{MeV}$ for all transitions below the
Planck scale.

{The following two points are in order concerning the role played by
the $X$ component. On the one hand, let us note that we have taken
$X$ as representing the full dark matter (DM) content of the
Universe. This is a possibility, which we have chosen for
definiteness in this presentation of the model, in part for
simplicity and also because, then, the quantity $\rD=\rX+\rLeff\ $
provides a kind of economical unification of the DM and DE parts
into an overall, self-conserved, dark sector. However, this ansatz
must be further elaborated. In particular, it must be confronted
with the structure formation data from the analysis of cosmic
perturbations in this model\,\cite{Relax02}. Another possibility
would be, for instance, that the ``conventional DM'' is contained in
what we have called the ordinary energy-momentum tensor in
Eq.\,(\ref{GEE}). In this alternative scenario, the total matter
content is covariantly conserved, and $X$ appears as a kind of
additional entity in the DE sector, which would interact with the
effective CC, and whose only purpose is to make the total DE
covariantly conserved. Since we still have $\omega_X=0$, the entity
$X$ looks now more as a new form of DM that is integrated into the
DE.

On the other hand, once the dynamics of $\rLeff$ is fixed, in this
case through (\ref{eq:rhoL-Intro}) and (\ref{eq:rho_L}), the
evolution of the component $X$, in whatever of the two options
discussed above, is completely determined by the local covariant
conservation law (\ref{eq:Bianchi}). This implies that $X$ cannot be
generally assimilated to a scalar field, because a dynamical scalar
field with some particular potential has its own dynamics. In this
sense, $X$ is to be viewed as an effective entity within the
generalized field equations. In the particular option that we have
analyzed in Fig.\,\ref{fig:EnergyDensities}, it is supposed to mimic
all effects associated to a real DM substratum.}

\begin{figure}[t]
\noindent\begin{centering}
\includegraphics[clip,width=1\columnwidth]{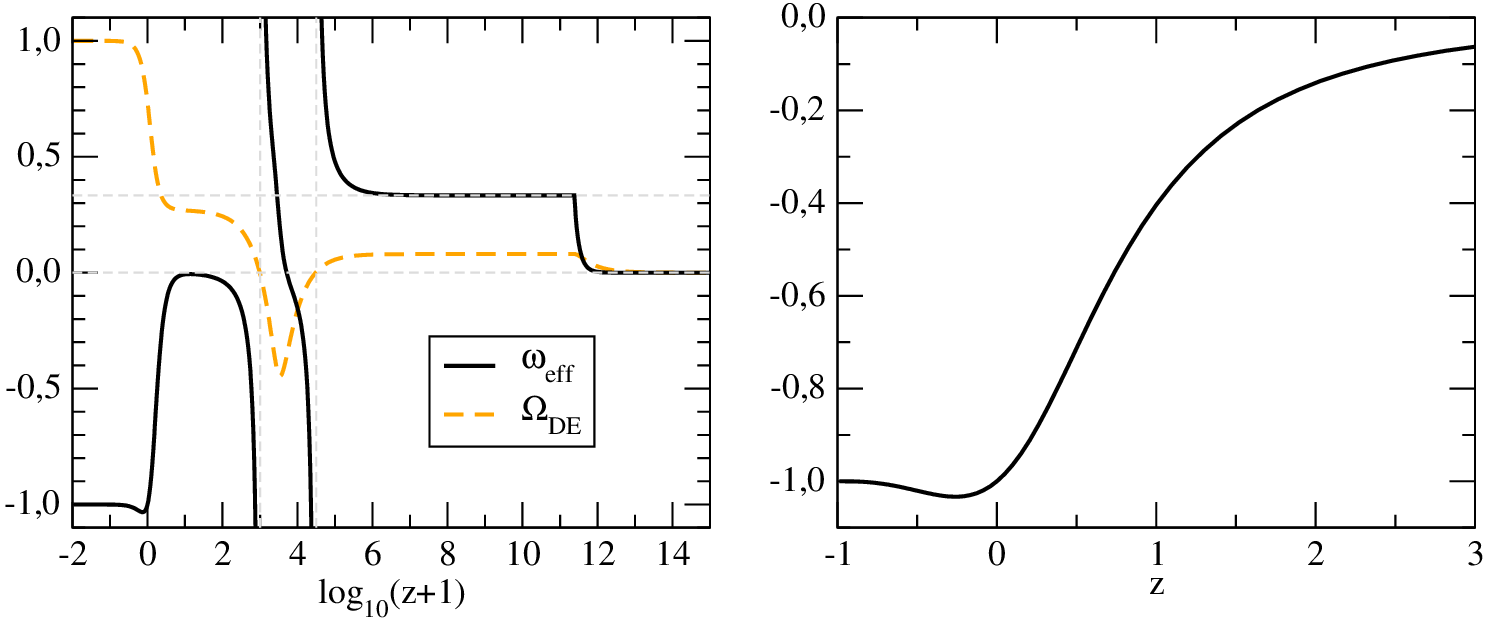}
\par\end{centering}
\caption{\label{fig:EffectiveEOS2plots}Effective equation of
state~$\omega_{\text{eff}}$ and relative effective DE
density~$\Omega_{\text{DE}}=\rho_{\text{DE}}/\rho_\text{tot}$. Parameters are as in Fig.~\ref{fig:EnergyDensities}.}
\end{figure}

\section{Discussion}\label{discussion}

Let us now further describe the different stages of cosmic evolution
in this framework. First, in the matter era, the total energy density is dominated by the DM component $\rX$ rather than by the vacuum energy~$\rLeff$. Due to $q\simeq{1}/{2}$ this can be understood easily by eliminating $\rX$ from Eqs.~(\ref{eq:Fried1},\ref{eq:Fried2}),
\begin{equation}
H^{2}\left(q-\frac{1}{2}\right)=
\frac{H_0^2}{\rco}\left(\frac{1}{2}\rR-\frac{3}{2}\rLeff\right)\ll H^2
\approx\frac{H_0^2}{\rco}\left(\rX+\rB\right).
\end{equation}
Therefore, this epoch behaves very similar to the $\Lambda$CDM matter era.
 Finally, the vacuum component becomes dominant at very late times and
the Universe smoothly enters the eternal de~Sitter regime with a
very small positive $\rLeff\simeq\rLo\ll\rLi$. In both eras, the relaxation model does not deviate much from the $\Lambda$CDM model, and the sign of the large initial vacuum density~$\rLi$ is not relevant.

Significant deviations from standard cosmology emerge in the
radiation era, because there is no constraint that enforces~$\rLeff$
to be negligible. In fact, the exact behavior of~$\rX$ and~$\rLeff$
depends on initial conditions and the details of~$f$. Nevertheless,
in that epoch (for which $q\simeq 1$ and $\rB$ is negligible),
subtraction of Eqs.~(\ref{eq:Fried1},\ref{eq:Fried2})  leads to
\begin{equation}
\frac{R}{6H_{0}^{2}}=\frac{H^{2}}{H_{0}^{2}}(1-q)=\frac{1}{\rho_{c}^{0}}\left(\frac{1}{2}\rho_{X}+2\rLeff\right).
\end{equation}
Moreover, the second term in the function~$f$ in
Eq.~(\ref{eq:rho_L}) is dominant in the radiation era (by
construction). Thus, while radiation dominates we have
\begin{equation}
 \rLeff=\rLi+\frac{\beta}{y\cdot72H^{6}(1-q)(q^{2}+1)}=\rLi+\frac{\beta}{y\cdot24H^{4}}\cdot\frac{1}{R},
\end{equation}
and by eliminating the Ricci scalar~$R$ from the two previous
equations, we obtain
\begin{equation}
 \rLeff= \rLi+\frac{\gamma}{(\frac{1}{2}\rho_{X}+2\rLeff)}.\label{eq:rhoL-Rad}
\end{equation}
With $\beta\sim-\rLi H_{0}^{2}$, $y=(H_{eq})^{-4}$ and the Hubble
rate~$H_{eq}\sim10^{5}H_{0}$ at the radiation--matter transition
we find that the variable
\begin{equation}
 \gamma=\rho_{c}^{0}\cdot\frac{\beta}{y\cdot24H^{4}\cdot6H_{0}^{2}}\approx \rho^{0}_{c}\cdot\frac{-\rLi}{144}\cdot\left(\frac{H_{eq}}{H}\right)^{4}
\end{equation}
becomes subdominant for very large Hubble rates~$H\gg H_{eq}$.  Eq.~(\ref{eq:rhoL-Rad}) has two solutions for~$\rLeff$,
\begin{equation}
 \rho_{\pm}=\frac{1}{8}\left(4\rLi-\rho_{X}\pm\sqrt{32\gamma+(4\rLi+\rho_X)^{2}}\right),
\end{equation}
and the physical one has to be compatible with $|\rLeff |\ll |\rLi|$ at late times. Also, $\rX$ is taken to be positive.

For negative initial vacuum energy~$\rLi<0$ the following limits
exist. At very high redshift, where $\rho_{X}\gg -4\rLi$, the
effective vacuum energy acts like a true (and subdominant) constant,
\begin{equation}
 \rho_{+}\simeq\frac{1}{8}\left(4\rLi-\rho_{X}+|4\rLi+\rho_{X}|\right)
 =\rLi.
\end{equation}
Thus, the universe evolves in a standard way, and the X component
redshifts like non-interacting dust. At smaller redshift, in the
range when $\sqrt{\gamma}\ll\rho_{X}\ll-4\rLi$ holds, we find a
(temporary) tracking regime:
\begin{equation}
 \rho_{+}\simeq\frac{1}{8}\left(4\rLi-\rho_{X}-(4\rLi+\rho_{X})\right)
 =-\frac{1}{4}\rho_{X}\,.
 \label{eq:rplus}
\end{equation}
In this regime, $\rLeff$ not only tracks the energy density of the X
component, but also that of radiation $\rR$. Indeed, in view of
(\ref{eq:rplus}), the conservation equation~(\ref{eq:Bianchi}) takes
on the form
\begin{equation}
\dot{\rho}_i+4H\rho_{i}\simeq 0\ \  ( \text{for both} \ \rho_i=\rX,
\rLeff )\label{eq:Bianchi2}
\end{equation}
during this regime.  Finally, for $\rho_{X}\ll\sqrt{\gamma},|\rLi |$ the
relaxation of the CC becomes obvious by
\begin{equation}
 \rho_{+} \simeq
 \frac{1}{8}\left(4\rLi-\rX +|4\rLi+\rX| +\frac{32\gamma}{2\cdot4|\rLi|}+\mathcal{O}((\rLi)^{-2})\right)
 \ll|\rLi|.
\end{equation}
Note that the analytical discussion is nicely supported by the
numerical results shown in Fig.~\ref{fig:EnergyDensities-highz}.

This relaxation regime also exists for positive initial $\rLi>0$,
when $\rho_{X}\ll\sqrt{|\gamma |},\rLi$,
\begin{equation}
 \rho_{-} \simeq
 \frac{1}{8}\left(4\rLi-\rX -(4\rLi+\rX) -\frac{32\gamma}{2\cdot4\rLi}+\mathcal{O}((\rLi)^{-2})\right)
 \ll\rLi.
\end{equation}
Whereas for higher redshift, when $\rho_{X}\gg\sqrt{|\gamma |}$, we
also find a tracking regime, $\rho_{-}\simeq -\frac{1}{4}\rho_{X}$.
However, differently from the $\rLi<0$ case, this tracking behavior
is persistent even for $\rho_{X}\gg \rLi$. Consequently, at the end
of reheating, the energy densities of radiation, dark matter and dark
energy could be of the same order of magnitude.

Because of the tracking relation $\rX\simeq-4\rLeff\propto \rR$, we
should care about bounds from nucleosynthesis. At that time
($z\lesssim 10^{9}$), we have in our example $\OD=\OL+\OX\simeq
3\OX/4\approx 0.08$ versus $\OR\simeq 0.9$ (cf.\  Fig.~\ref{fig:EnergyDensities}), and so the ratio $\OD/\OR\lesssim
10\%$ is safe for standard Big Bang nucleosynthesis (similarly as
in\,\cite{LXCDM}). On the other hand, this model offers also the
possibility to solve the coincidence problem, in that $\rX$,~$\rR$
and $\rLeff$ are not very different during the tracking regime until
the matter-radiation transition (cf.
Fig.~\ref{fig:EnergyDensities-highz}). Note also that in comparison
to $\Lambda$CDM, much more dark matter is allowed before the
relaxation regime thereby weakening the constraints on dark matter
properties.

Finally, we discuss the effective EOS~$\omega_{\text{eff}}$, which is a useful tool
for comparing interacting DE models with non-interacting ones.
According to\,\cite{SS12}, $\omega_{\text{eff}}$ is given by the EOS
of a self-conserved DE component $\rho_{\text{DE}}$ in an universe
with the same Hubble rate~$H(z)$ and total energy
density~$\rho_{\text{tot}}$ as the relaxation model. Within this
effective description, DM obeys the usual scaling law of
dust~$\tilde{\rho}_X\propto(z+1)^{3}$ since the interaction with DE
is absent. Thus,
\begin{equation}\label{effEOS}
\omega_{\text{eff}}=-1+\frac{1+z}{3}\frac{1}{\rho_{\text{DE}}(z)}\frac{d\rho_{\text{DE}}(z)}{dz},
\end{equation}
with~$\rho_{\text{DE}}=\rho_{\text{tot}}-\tilde{\rho}_X-\rho_r-\rho_b$.
Since~$\omega_\text{eff}$ is more accessible for observations at low
redshift, we magnify this range in the second plot of
Fig.~\ref{fig:EffectiveEOS2plots}. In the relaxation regime,
$\omega_\text{eff}$ follows mostly the EOS of the dominant
component. Notice that $\Omega_{\text{DE}}$ is well defined
everywhere despite the behavior of $\omega_{\text{eff}}$.

\section{Conclusions}\label{conclusions}

In this Letter, we have addressed the old cosmological constant
problem, i.e.\ the difficult problem of relaxing the value of the
cosmological vacuum energy. The necessity of tempering this value
ably and plausibly is absolutely crucial for a realistic
cosmological evolution from the early times till today. Indeed, the
vacuum energy of the early Universe is expected to be huge in
Particle Physics standards, since the expansion history must drive
through a series of phase transitions of diverse nature; in
particular, it goes through a process of fast inflation (presumably
associated to some Grand Unified Theory) and also through the
spontaneous breaking of the electroweak symmetry. Finally, it
undergoes the more modest chiral symmetry breaking transition, which
occurs at the milder scale $\Lambda_{QCD}={\cal O}(0.1)$ GeV and is
connected to the quark-gluon-hadron transition. Even if the latter
would have been the only phase transition ever occurred, and the
associated vacuum energy density would still persist, it would be a
disaster for our Universe. The reason is that the value of the
cosmological constant associated to that energy density would have
accelerated the expansion history to the point of preventing the
formation of any of the structures that we now see in our cosmos, as
they would have been ripped off by the fast expansion rate during
the first stages of formation. However, being the Standard Model
(SM) of Particle Physics such an extraordinary successful
theoretical and experimental framework, we must conclude that both
of its important contributions to the vacuum energy (associated to
the electroweak and strong interactions) must have been relaxed very
fast after the corresponding phase transitions occurred; namely,
sufficiently fast as to insure not only the possibility to form
structures in the late Universe, but also to leave the
nucleosynthesis process fully unscathed after the first minute of
expansion.

In this Letter, we have tackled a possible cure to this longstanding
problem; we have proposed a dynamical relaxation mechanism of the
vacuum energy that operates at the level of the generalized
gravitational field equations. Our relaxation mechanism achieves
this goal without fine-tuning. Apart from the ordinary baryonic
matter, which in our model stays covariantly conserved, we assume
that the dark sector is made out of an effective cosmological term
$\rLeff$ and another dynamical component $X$ exchanging energy with
it. The cosmological term of our model is actually an effective one,
in the sense that it is defined as the sum of an arbitrarily large
cosmological constant, $\rLi$, and a particular combination of
curvature invariants, $\rinv$, such that the sum behaves as an
overall CC term $\rLeff(t)=\rLi+\rinv(t)$, but one that evolves with
time. As we have said, the component $X$ interacts with the variable
$\rLeff$, but the total density of the dark sector,
$\rD=\rX+\rLeff$, is covariantly conserved. This construction fits
into the class of the so-called $\CC$XCDM models existing in the
literature\,\cite{LXCDM}. Interestingly enough, in the present
context, the $X$ component can be interpreted as the dark matter
(DM) and the contribution $\rinv$ can be viewed as a (dynamical)
dark energy component that adds up to the traditional cosmological
constant term. However, such dynamical DE component has nothing to
do with scalar fields as it is of purely gravitational origin.
Therefore, the total energy density of the dark sector splits into
the sum of the various components DM, CC and DE, namely
$\rD=\rX+\rLi+\rinv$, whereas the ordinary (baryonic) matter does
not interact at all with the dark sector and remains safely
conserved.

The evolution of this $\CC$XCDM universe keeps the total DE
sub-dominant during the radiation and matter epochs, and only at
late times it leads to a tiny effective CC whose smallness is a
direct consequence of the large initial vacuum energy, $\rLi$,
rather than to a severe fine-tuning involving ugly cancelation of
large terms. The resulting cosmos can transit from a fast early
inflationary Universe, then drive through the standard radiation and
matter dominated epochs and, eventually, ends up in an extremely
slow de Sitter phase; in fact, a model of Universe very close to the
one suggested by the modern cosmological data\,\cite{cosmdata}.
Finally, we obtained new insights into the coincidence problem, as
we observed an interesting tracking behavior in the radiation era. A
more detailed exposition of our approach discussing the universality
of the CC relaxation and the corresponding analysis of the
cosmological perturbations will be presented
elsewhere\,\cite{Relax02}.



\subsection*{Acknowledgments}
The authors have been supported in part by DURSI Generalitat de Catalunya under project 2005SGR00564; FB and JS 
also by MEC and FEDER under project FPA2007-66665 and by the
Consolider-Ingenio 2010 program CPAN CSD2007-00042, and HS by the
Ministry of Education, Science and Sports of the Republic of Croatia
under contract No. 098-0982930-2864.

\newcommand{\JHEP}[3]{ {JHEP} {#1} (#2)  {#3}}
\newcommand{\NPB}[3]{{\sl Nucl. Phys. } {\bf B#1} (#2)  {#3}}
\newcommand{\NPPS}[3]{{\sl Nucl. Phys. Proc. Supp. } {\bf #1} (#2)  {#3}}
\newcommand{\PRD}[3]{{\sl Phys. Rev. } {\bf D#1} (#2)   {#3}}
\newcommand{\PLB}[3]{{\sl Phys. Lett. } {\bf B#1} (#2)  {#3}}
\newcommand{\EPJ}[3]{{\sl Eur. Phys. J } {\bf C#1} (#2)  {#3}}
\newcommand{\PR}[3]{{\sl Phys. Rep. } {\bf #1} (#2)  {#3}}
\newcommand{\RMP}[3]{{\sl Rev. Mod. Phys. } {\bf #1} (#2)  {#3}}
\newcommand{\IJMP}[3]{{\sl Int. J. of Mod. Phys. } {\bf #1} (#2)  {#3}}
\newcommand{\PRL}[3]{{\sl Phys. Rev. Lett. } {\bf #1} (#2) {#3}}
\newcommand{\ZFP}[3]{{\sl Zeitsch. f. Physik } {\bf C#1} (#2)  {#3}}
\newcommand{\MPLA}[3]{{\sl Mod. Phys. Lett. } {\bf A#1} (#2) {#3}}
\newcommand{\CQG}[3]{{\sl Class. Quant. Grav. } {\bf #1} (#2) {#3}}
\newcommand{\JCAP}[3]{{ JCAP} {\bf#1} (#2)  {#3}}
\newcommand{\APJ}[3]{{\sl Astrophys. J. } {\bf #1} (#2)  {#3}}
\newcommand{\AMJ}[3]{{\sl Astronom. J. } {\bf #1} (#2)  {#3}}
\newcommand{\APP}[3]{{\sl Astropart. Phys. } {\bf #1} (#2)  {#3}}
\newcommand{\AAP}[3]{{\sl Astron. Astrophys. } {\bf #1} (#2)  {#3}}
\newcommand{\MNRAS}[3]{{\sl Mon. Not. Roy. Astron. Soc.} {\bf #1} (#2)  {#3}}
\newcommand{\JPA}[3]{{\sl J. Phys. A: Math. Theor.} {\bf #1} (#2)  {#3}}
\newcommand{\ProgS}[3]{{\sl Prog. Theor. Phys. Supp.} {\bf #1} (#2)  {#3}}
\newcommand{\APJS}[3]{{\sl Astrophys. J. Suppl.} {\bf #1} (#2)  {#3}}

\newcommand{\Prog}[3]{{\sl Prog. Theor. Phys.} {\bf #1}  (#2) {#3}}
\newcommand{\IJMPA}[3]{{\sl Int. J. of Mod. Phys. A} {\bf #1}  {(#2)} {#3}}
\newcommand{\IJMPD}[3]{{\sl Int. J. of Mod. Phys. D} {\bf #1}  {(#2)} {#3}}
\newcommand{\GRG}[3]{{\sl Gen. Rel. Grav.} {\bf #1}  {(#2)} {#3}}




\end{document}